\documentclass[useAMS,usenatbib]{mn2e}
\usepackage[dvips]{graphicx,color}

\usepackage{amssymb}
\usepackage{indentfirst}
\usepackage{epsfig}
\usepackage{pdflscape}
\usepackage{afterpage}

\def\nus{{\it NuSTAR }}
\def\sw{{\it Swift}}
\def\fm{{\it Fermi}}
\def\HXMT{{\it HXMT }}
\def\nic{{\it Nicer }}

\begin{document}
\title[]{Comparing the Super-Eddington accretion of SMC X-3 and RX J0209.6-7427
with Swift J0243.6+6124}

\author[J. Liu et al.]{Jiren Liu$^{1}$\thanks{E-mail: liujiren@bjp.org.cn},
Georgios Vasilopoulos$^2$, MingYu Ge$^3$, Long Ji$^4$, Shan-Shan Weng$^5$, 
\newauthor
Shuang-Nan Zhang$^3$, Xian Hou$^{6,7}$\\
$^{1}$Beijing Planetarium, 138 Sizhimenwai Road, Beijing 100044, China\\
$^{2}$Observatoire astronomique de Strasbourg, UMR 7550, F-67000 Strasbourg, France  \\
$^{3}$Institute of High Energy Physics, Chinese Academy of Sciences, Beijing 100049, China\\
$^{4}$School of Physics and Astronomy, Sun Yat-sen University, 2 Daxue Road, Zhuhai, Guangdong 519082, China\\
$^{5}$Department of Physics and Institute of Theoretical Physics, Nanjing Normal University, Nanjing 210023, China\\
$^{6}$Yunnan Observatories, Chinese Academy of Sciences, Kunming 650216, China \\
$^{7}$Key Laboratory for the Structure and Evolution of Celestial Objects, Chinese Academy of Sciences, Kunming 650216, China
 }

\date{}

\maketitle

\begin{abstract}
We study the giant outbursts of SMC X-3 and RX J0209.6-7427 
to compare their super-Eddington 
accretion regime with that of Swift J0243.6+6124. 
The high double-peak profile of SMC X-3 
is found to be 0.25 phase offset from that below $2.3\times10^{38}$erg\,s$^{-1}$, 
which is similar to Swift J0243 (happened around $0.9\times10^{38}$erg\,s$^{-1}$). 
The profile of RX J0209 shows a similar 0.25 phase offset between high double-peak and
low double-peak around $1.25\times10^{38}$erg\,s$^{-1}$. 
The 0.25 phase offset corresponds to a 90 degree angle change of the emission beam
and strongly supports for a 
transition from a fan beam to a pencil beam.
Their critical luminosities imply a surface magnetic field $\sim4\times10^{13}$ G and $2\times10^{13}$ G for SMC X-3 and RX J0209, respectively, based on the recently measured cyclotron line of Swift J0243.
The spin-up rate and luminosity of SMC X-3 
follows a relation of $\dot{\nu}\propto L^{0.94\pm0.03}$, while 
that of RX J0209 follows $\dot{\nu}\propto L^{1.00\pm0.03}$, which are similar to 
Swift J0243 and consistent 
with the prediction of a radiation-pressure-dominated (RPD) disk. 
These results indicate that accretion columns are indeed 
formed above Eddington luminosity, and the population of ULXPs likely 
corresponds to X-ray pulsars of highest magnetic field.
\end{abstract}


\section{Introduction}

The discovery of ultra-luminous X-ray pulsars (ULXPs)
revealed
the existence of super-Eddington accretion onto magnetized neutron stars 
with luminosity around $10^{40}$erg\,s$^{-1}$ \citep[e.g.][]{Bac14,Fur16,Isr17a,Isr17b,Car18}.
At such a high luminosity, an accretion column will be formed below 
a radiation shock around the magnetic pole. The accretion disk is 
most likely geometrically thick,
and an outflow driven by the strong radiation field is expected \citep{SS73}.
The process of super-Eddington accretion is relatively poorly studied compared with the 
sub-Eddington accretion, due to the rareness of targets and limited availability of 
good data since most ULXPs are located in other galaxies 
\citep[for a recent review on the topic, please refer to][]{Fab21, MT22}.

The discovery of the Galactic ULXP of Swift J0243.6+6124 (J0243 hereafter) during
its 2017
giant outburst, which reached a luminosity above $10^{39}$erg\,s$^{-1}$,
has enabled a detailed study of super-Eddington process around its outburst 
peak\citep[e.g.][]{Tsy18,Wil18,Tao19,Zhang19,Dor20}.
The detailed temporal study \citep{Liu22} showed that the formation of 
accretion column was evidenced by a 0.25 phase offset, 
between the double-peak profiles at luminosities below 
$1.5\times10^{38}$erg\,s$^{-1}$ and above $5\times10^{38}$erg\,s$^{-1}$.
The 0.25 phase offset corresponds to a 90 degree angle change of the 
emission beam, which is just the 
transition angle between a pencil beam and a fan beam.
The luminosity range within $1.5-5\times10^{38}$erg\,s$^{-1}$ corresponds to
an intermediate formation stage of accretion column from appearance 
to fully dominating the emission beam.

\begin{figure*}
	\includegraphics[width=3.4in]{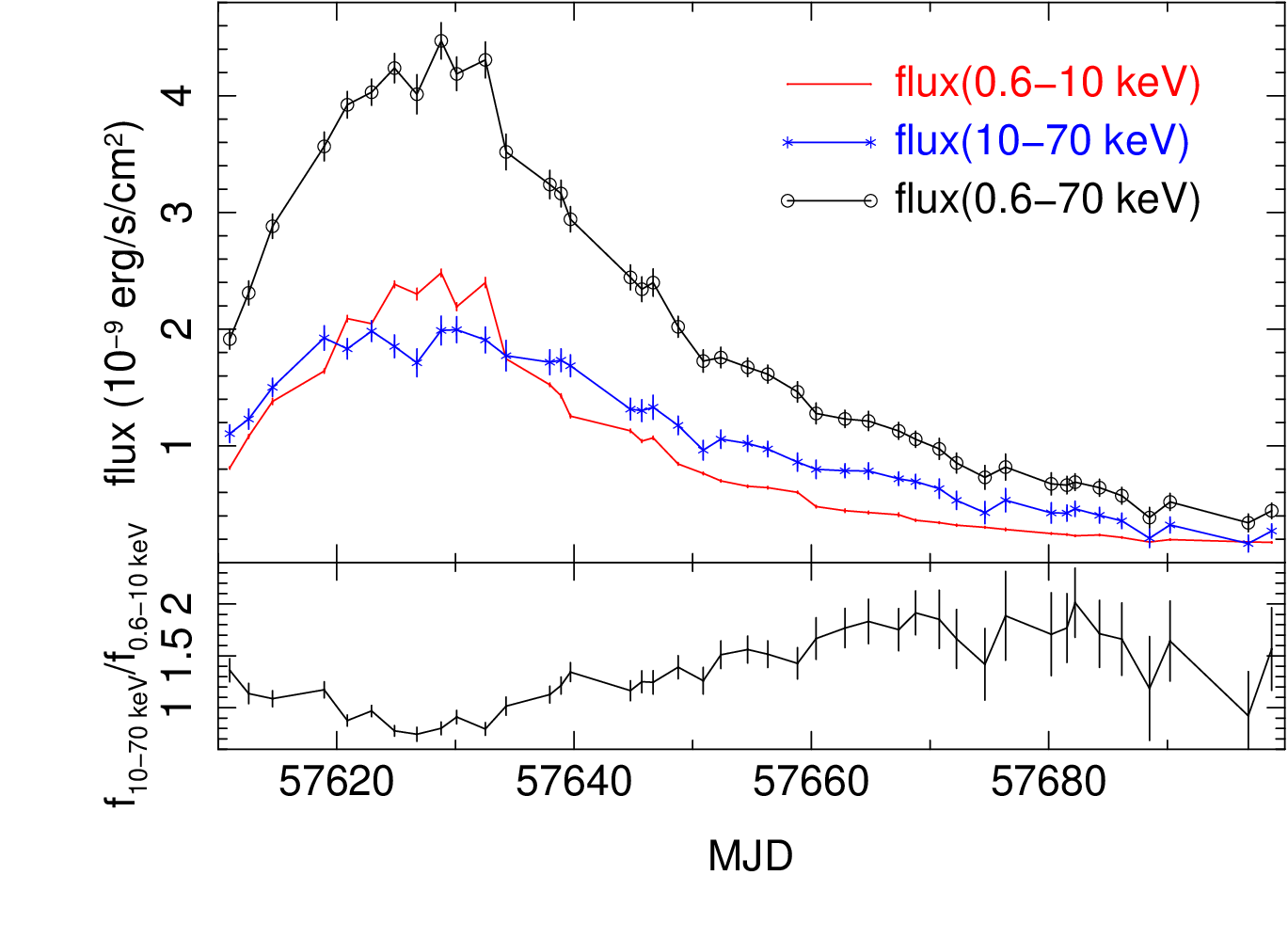}
	\includegraphics[width=3.4in]{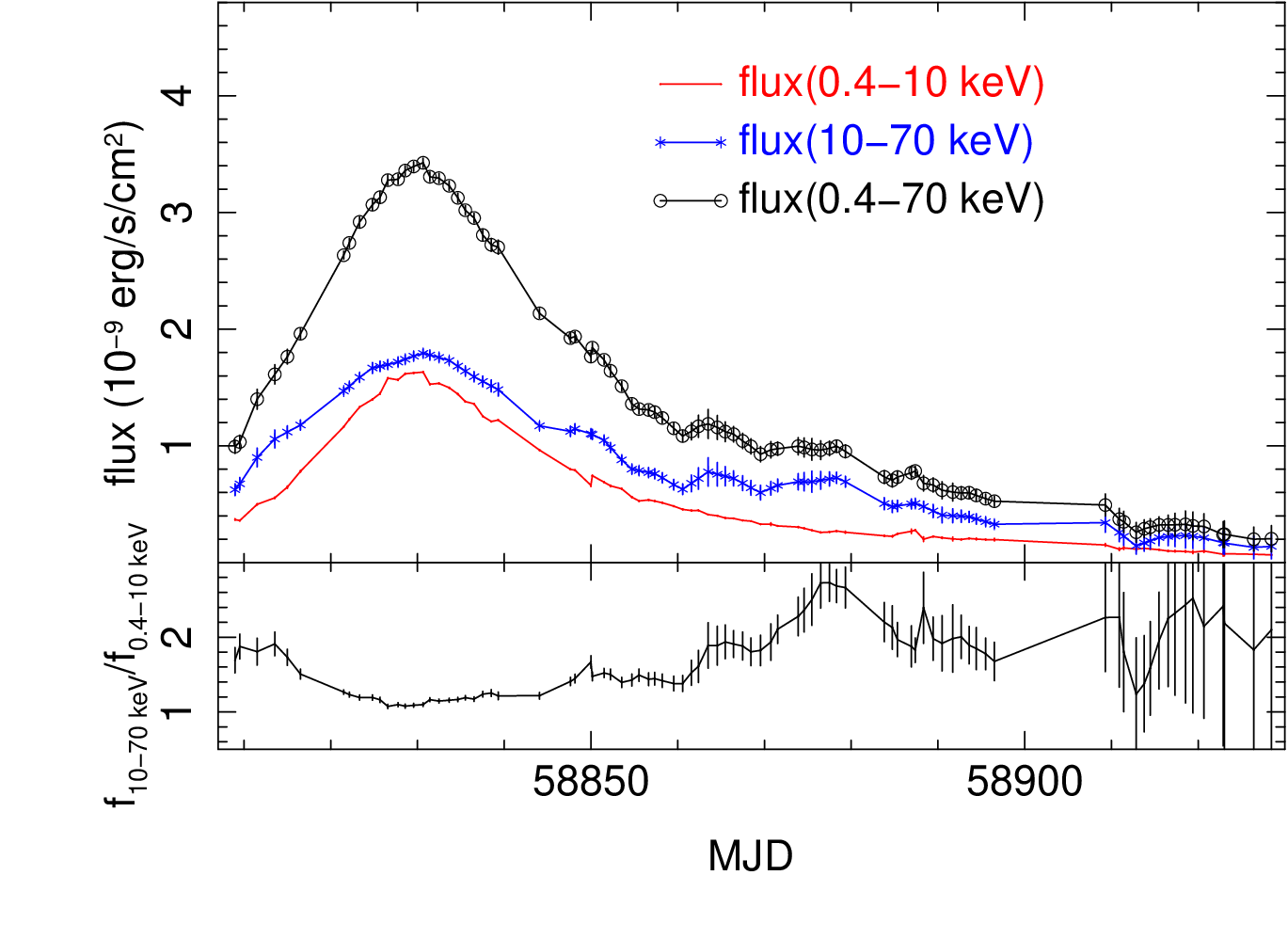}
	\caption{Bolometric fluxes of SMC X-3 (left) and RX J0209 (right), together 
with the fluxes in the soft and hard band and their ratio. The soft fluxes
	of SMC X-3 are from Swift/XRT, and those of RX J0209 are from {\it Nicer},
	while the hard fluxes are from Swift/BAT. 
	} 
\end{figure*}

\begin{figure*}
	\includegraphics[width=3.4in]{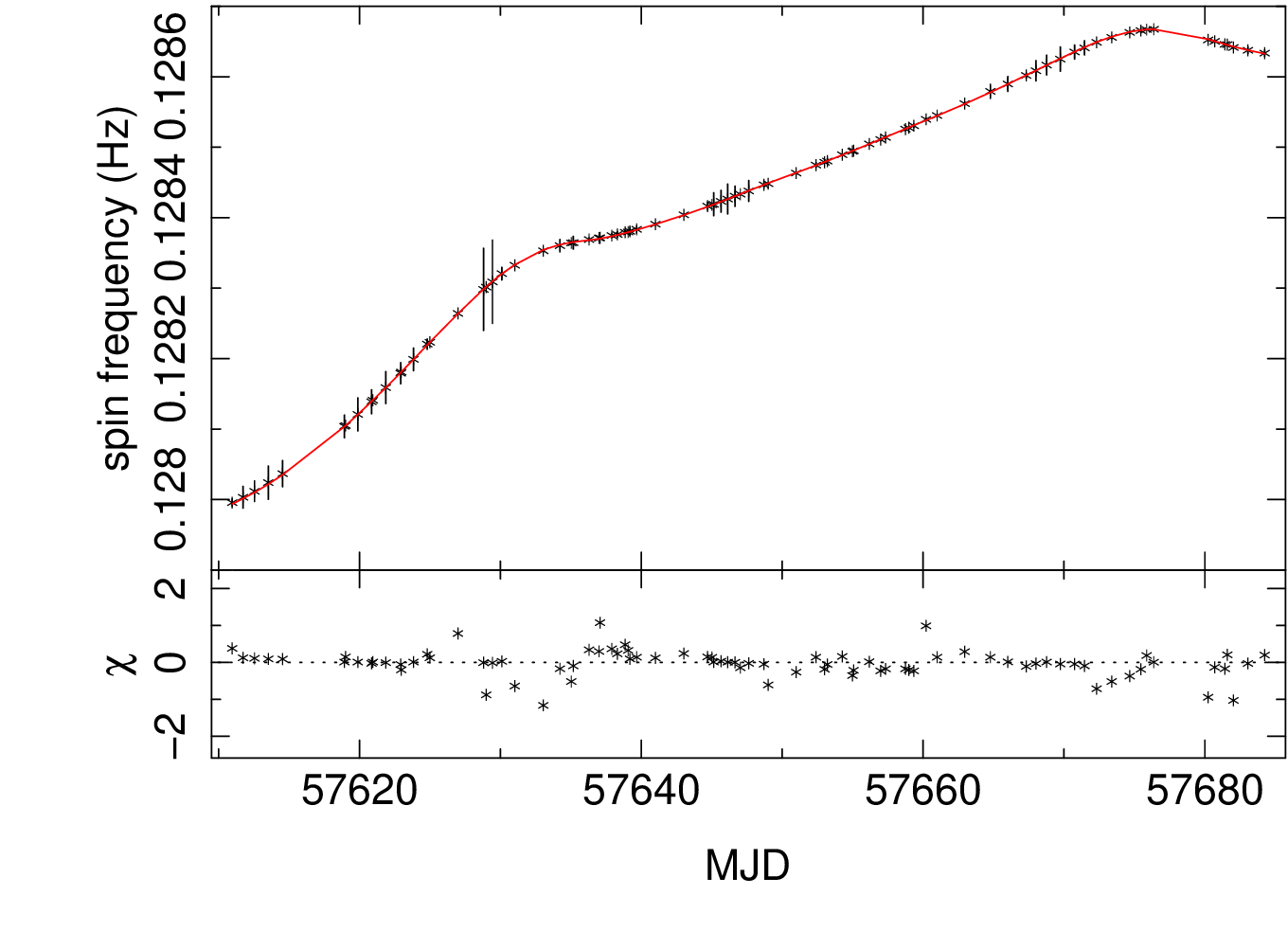}
	\includegraphics[width=3.4in]{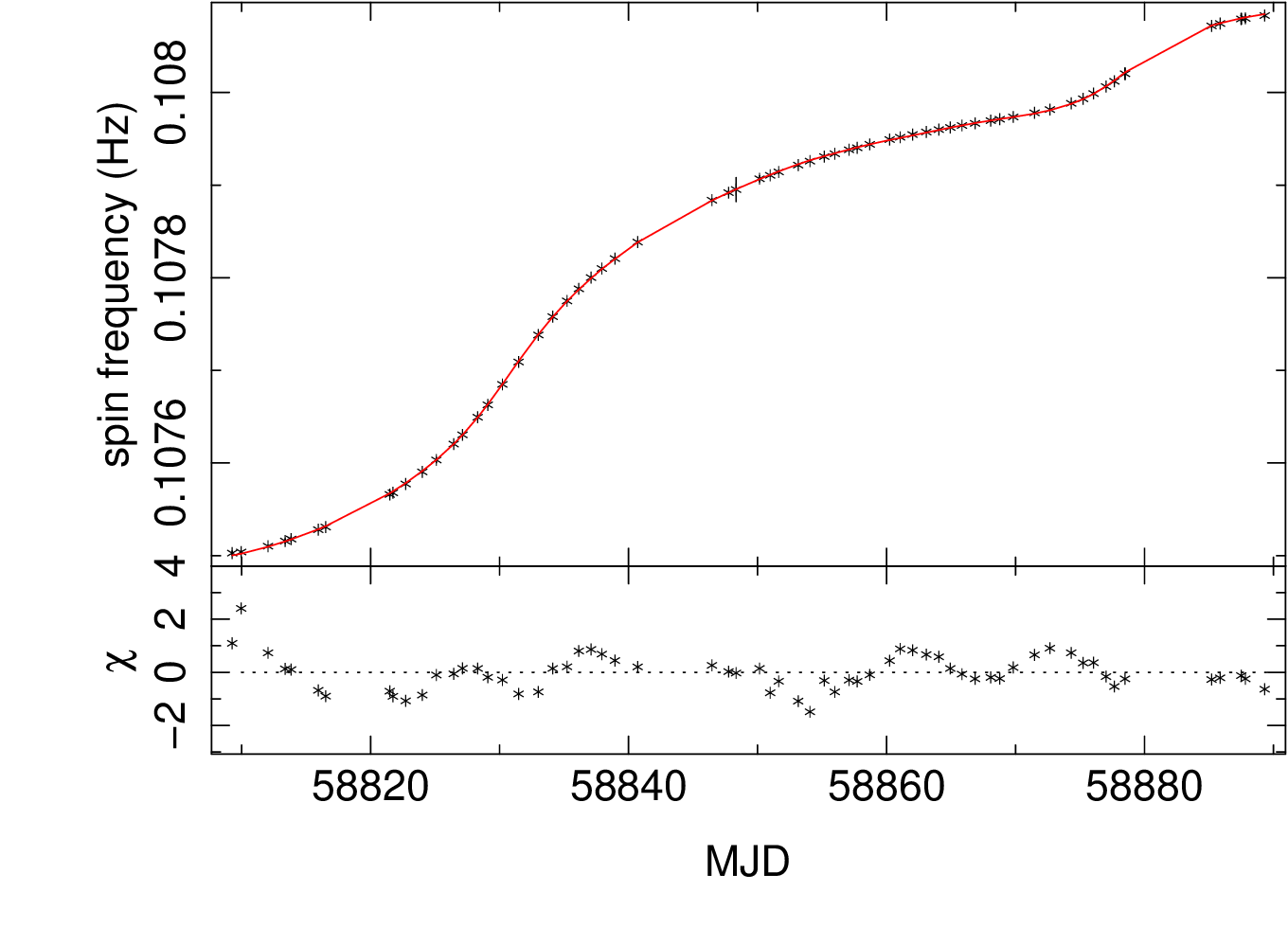}
	\caption{The barycenter-corrected spin frequencies of SMC X-3 (left) and 
RX J0209 (right), together with 
	the fitted value from a binary orbit plus pulsar torque model.
	} 
\end{figure*}

\begin{table*}
\scriptsize
\begin{center}
\caption{Fitted orbital parameters}
\label{tab2}
\begin{tabular}{ccc|cc}
\\ \hline
	parameters & This work & GBM team$^a$ & This work & GBM team$^a$ \\   
	 & SMC X-3 & SMC X-3 & RX J0209 & RX J0209 \\   \hline
	Orbital period (days) &  $45.2\pm0.2$       & 45.383 & $47.8\pm0.3$&  47.4 \\
	Time of periastron (MJD) & $57677.2\pm0.3$ & 57677  &$58782.3\pm0.5$ & 58782.56 \\
	asin$i$ (light-sec) & $195.2\pm1.5$        & 196.2     &$132.3\pm2.4$ & 169.8\\
	Long. of periastron (deg) & $208.0\pm2.5$   & 208.77      &$71.5\pm1.5$ & 65.7 \\   
	eccentricity   & $0.26\pm0.01$             &  0.22     &$0.36\pm0.02$ & 0.32 \\ 
	$\alpha$   & $0.95\pm0.02$    & -   & $0.97\pm0.02$ & -   \\ \hline

\end{tabular}
\begin{description}
         \begin{footnotesize}
         \item
            $^a$ https://gammaray.msfc.nasa.gov/gbm/science/pulsars
         \end{footnotesize}
   \end{description}

\end{center}
\end{table*}

A detailed model of radiation-pressure-dominated (RPD) disk \citep{Cha17,Cha19}
indicated that an increasing accretion rate
leads to an increased disk thickness, and the pressure balance 
can be satisfied at a similar radius for different accretion rate. 
As a result, the magnetosphere size of a RPD disk
is almost independent on the mass accretion rate and 
a linear-like relation of $\dot{\nu}\propto\dot{m}$ is expected. 
Indeed, the spin-up rate $\dot{\nu}$ of Swift J0243 due to accreted material 
is found to be linearly correlated with luminosity $L$ within 
$5-13\times10^{38}$erg\,s$^{-1}$ \citep{Liu22}.
However, above $1.3\times10^{39}$erg\,s$^{-1}$,
the $\dot{\nu}-L$ relation of Swift J0243 becomes flattened, 
coincident with a flattening
of spectral hardness and a saturation of the faint peak of the pulse profile.
These anomalous behaviors above $1.3\times10^{39}$erg\,s$^{-1}$
could be related with a geometry change of accretion 
disk due to the illumination by the strong radiation of accretion
column \citep{Liu22}. 
We note that the luminosity of Swift J0243 mentioned above would be
reduced by a factor of 0.6 if adopting a distance of 5.2 kpc measured 
from {\it Gaia} data 3 (see \S 4).

It is interesting to check the temporal behavior of 
other ULXP to see whether the behavior of Swift J0243 is a general feature of 
super-Eddington process. In this paper, we study the two ULXPs around Small Magellanic 
Cloud, SMC X-3 and RX J0209.6-7427 (hereafter J0209), the giant outbursts of 
which reached a peak luminosity around $1-2\times10^{39}$erg\,s$^{-1}$,
and they are close enough to allow detailed studies. 

The 2016 giant outburst of SMC X-3 was initially reported by MAXI on 
August 8, 2016 (MJD 57609) \citep{Neg16} and 
was identified by Swift/XRT \citep{Ken16}. It was regularly monitored 
by Swift/XRT, the data of which 
had been reported in the literature \citep{Weng17, Tow17,Tsy17}. 
Two \nus observations and one XMM-Newton observation were also performed
for SMC X-3 during the outburst \citep{Tsy17, Zhao18, KV18}.
The 2019 giant outburst of RX J0209 was detected by MAXI on 
November 20, 2019 (MJD 58807) \citep{Neg19} and was localized 
by Swift/XRT \citep{Ken19}, and it was regularly monitored by \nic
\citep{Iwa19,Vas20}. It was also observed once by \nus during the rising stage 
and several times by {\it Astrosat} \citep{Cha20} and 
{\it Insight-HXMT} \citep{Hou22} during the peak stage.
A distance of 62 kpc and 55 kpc is adopted for SMC X-3 and RX J0209, respectively.

\section{Light curves and orbital parameters}
\subsection{SMC X-3}

The binary orbital motion of SMC X-3 has an apparent effect on the observed 
spin frequency and has to be taken into account to obtain the intrinsic 
spin frequency. A classical relation of accreting torque of 
$\dot{\nu}\propto L^{6/7}$ was generally assumed when fitting the orbital solution.
\citet{Tow17} had found that such a relation can not provide a reasonable fit
to the observed spin frequency of SMC X-3. They noted that 
if the super-Eddington part of the giant outburst is excluded, a much improved 
fit is resulted. 
If the super-Eddington accretion of the giant outburst of SMC X-3 is similar 
to that of Swift J0243, a linear-like relation of $\dot{\nu}\propto L$ is
expected during the peak period, 
and the general usage of $L^{6/7}$ relation would be improper. 

On the other hand, \citet{Tow17} used the luminosity only within the XRT band
(0.5-10 keV), and 
\citet{Tsy17} noted that if correcting the 0.5-10 keV luminosity with a 
bolometric factor, 
they could obtain a much better fit. Therefore, to obtain a reasonable fitting 
of the orbital parameters, a bolometric luminosity and a limited time
range of the giant outburst period should be used, since the 
$\dot{\nu}-L$ relation is expected to change at different accretion luminosity.

\citet{Tsy17} estimated the bolometric correction of SMC X-3 based on 
the flux ratio in the 0.5-10 keV (Swift/XRT) plus 15-50 keV 
(Swift/BAT) bands to that in the 0.5-10 keV band, which is scaled to 
match two \nus observations (on MJD 57613 and 57704). 
We note that when converting the Swift/BAT rate to flux, 
they adopted a default converting factor suitable only for a power-law model with 
index of $\Gamma=2.15$ \citep[for Crab,][]{Kri13}, which is quite different from the index of SMC X-3
($\Gamma\sim1$). 
So we estimate a broadband flux as the sum of the flux in 
0.6-10 keV and in 10-70 keV bands. 
To estimate the unabsorbed flux in 
0.6-10 keV, we fit the XRT spectrum with a model of power-law plus black-body subjecting 
to a fixed absorption column of $0.14\times10^{22}$ cm$^{-2}$ \citep{Zhao18} and infer the 
unabsorbed flux by setting the model absorption to zero.

To estimate the flux in 
10-70 keV band, we convert the BAT 15-50 keV rate to 10-70 keV 
flux based on \nus spectra.
The \nus flux on MJD 57613 is $1.41\times10^{-9}$erg\,cm$^{-2}$\,s$^{-1}$
within 10-70 keV band, while 
the BAT rate at this time is about 0.0149 counts\,cm$^{-2}$\,s$^{-1}$ ($C_1$) estimated from a
two-days-binned BAT light curve. So a converting factor of 
$f_1=1.41\times10^{-9}/0.0149=0.95\times10^{-7}$ erg/counts is resulted. 
This is 1.65 times larger than the default converting factor mentioned above 
($1.26\times10^{-8}/0.22=0.573\times10^{-7}$ erg/counts).
For the second \nus observation on MJD 57705, the \nus flux and BAT 
rate are $2.48\times10^{-10}$erg\,cm$^{-2}$\,s$^{-1}$ 
and 0.0026 counts\,cm$^{-2}$\,s$^{-1}$ ($C_2$), respectively, and a factor 
of $f_2=0.96\times10^{-7}$ erg/counts is resulted.
As the rate is relatively low on MJD 57705,
we adopt the factor of $0.95\times10^{-7}$ erg/counts
to convert the BAT rate to the flux within 10-70 keV.

The obtained bolometric fluxes, together with those in 0.6-10 keV (XRT) 
and 10-70 keV (BAT) bands and their ratio, are plotted in Figure 1.
The soft band fluxes are a little higher than those in the hard band
during the outburst peak between MJD 57620 and 57633, while during other time, 
the soft band fluxes were lower than the hard band fluxes. This reflects
the softening of the spectrum near the outburst peak.
The hard-to-soft ratio reached a level of 2 around MJD 57670-57680.

The spin frequencies of SMC X-3 had been measured with
both Fermi/GBM and Swift/XRT data.
In principle, the GBM measurements are more accurate as the 
GBM daily coverage is longer than those for Swift/XRT, which generally
has only a few snapshots for one observation. 
Nevertheless, the XRT spin measurements are helpful supplement
because the regular GBM measurements only extended to MJD 57673.
To measure the spin frequencies from barycenter-corrected XRT light curve,
we first estimate a spin period for one XRT observation with epoch folding method,
and then we employ a phase connection between the snapshots 
to obtain a more accurate estimation.

To fit the orbital parameters, 
we adopt a torque model of $\dot{\nu}=bL_{bol}^\alpha$.
We use both the GBM and XRT barycenter-corrected spin frequency measurements.
For the GBM data, we neglected those data before MJD 57616, which showed some
fluctuations. 
We limit the fitting time range to MJD 57610-57685, which includes the 
outburst peak and two turn-over of measured spin frequencies. 
The fitting is performed for frequency vs time with $L_{bol}$ as input.
The fitted results are listed in Table 1 and plotted in Figure 2.
In general, the fitted orbital parameters are consistent with those reported 
in previous studies. Compared with that obtained by Fermi/GBM pulsar team, the most 
different parameter is the eccentricity: we got $e=0.26\pm0.1$, while 
their value is 0.22. An eccentricity of $e=0.26$ is also 
obtained by \citet{Weng17}, who used a high-order polynomial 
to follow the evolution of the intrinsic spin instead of a torque model.
We note that a different $e$ will change the intrinsic $\dot{\nu}$ apparently 
near periastron.
The fitted index of $\alpha=0.95\pm0.02$ is larger than the 
standard index of 6/7, similar as Swift J0243. 

\begin{figure*}
	\includegraphics[width=3.4in]{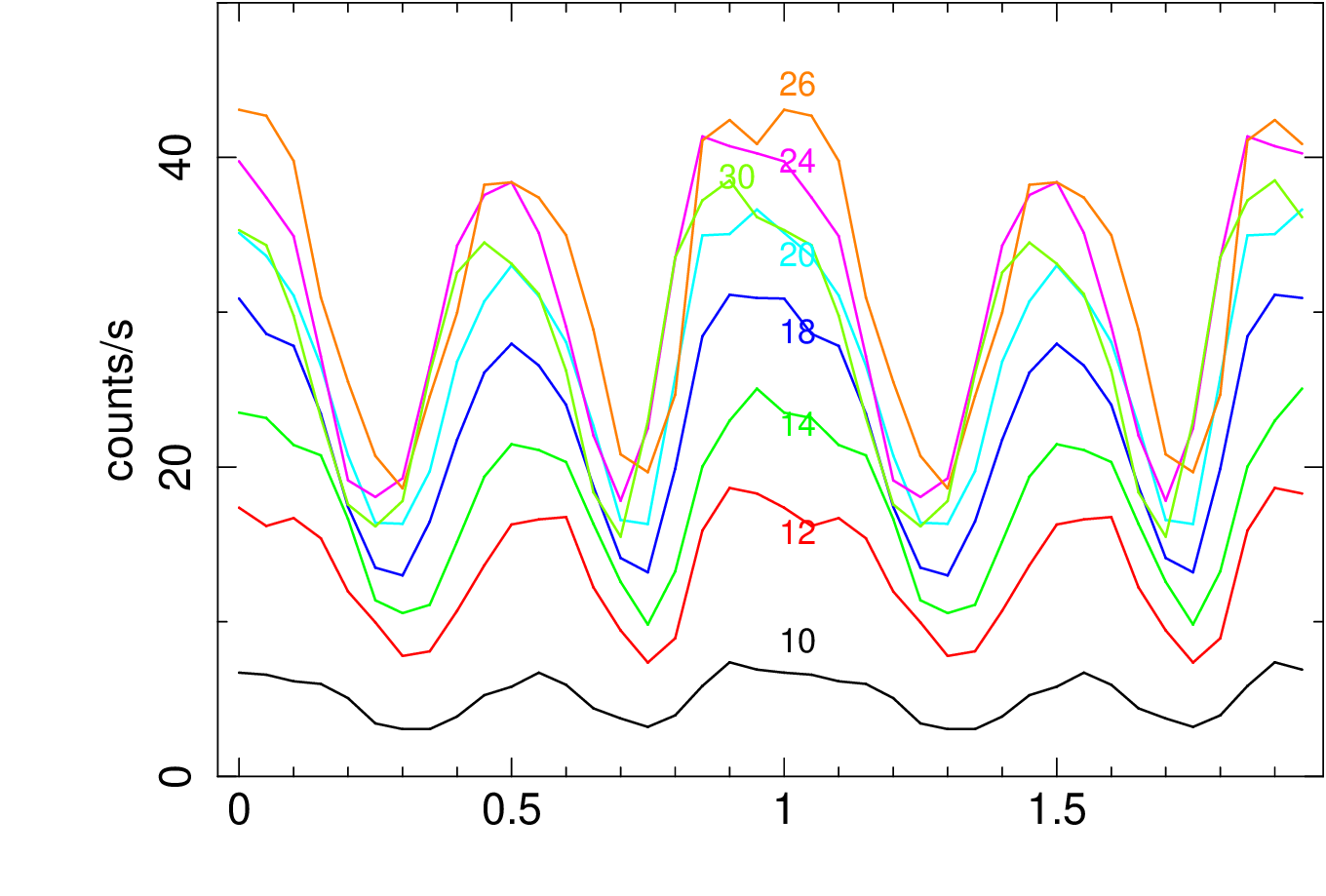}
	\includegraphics[width=3.4in]{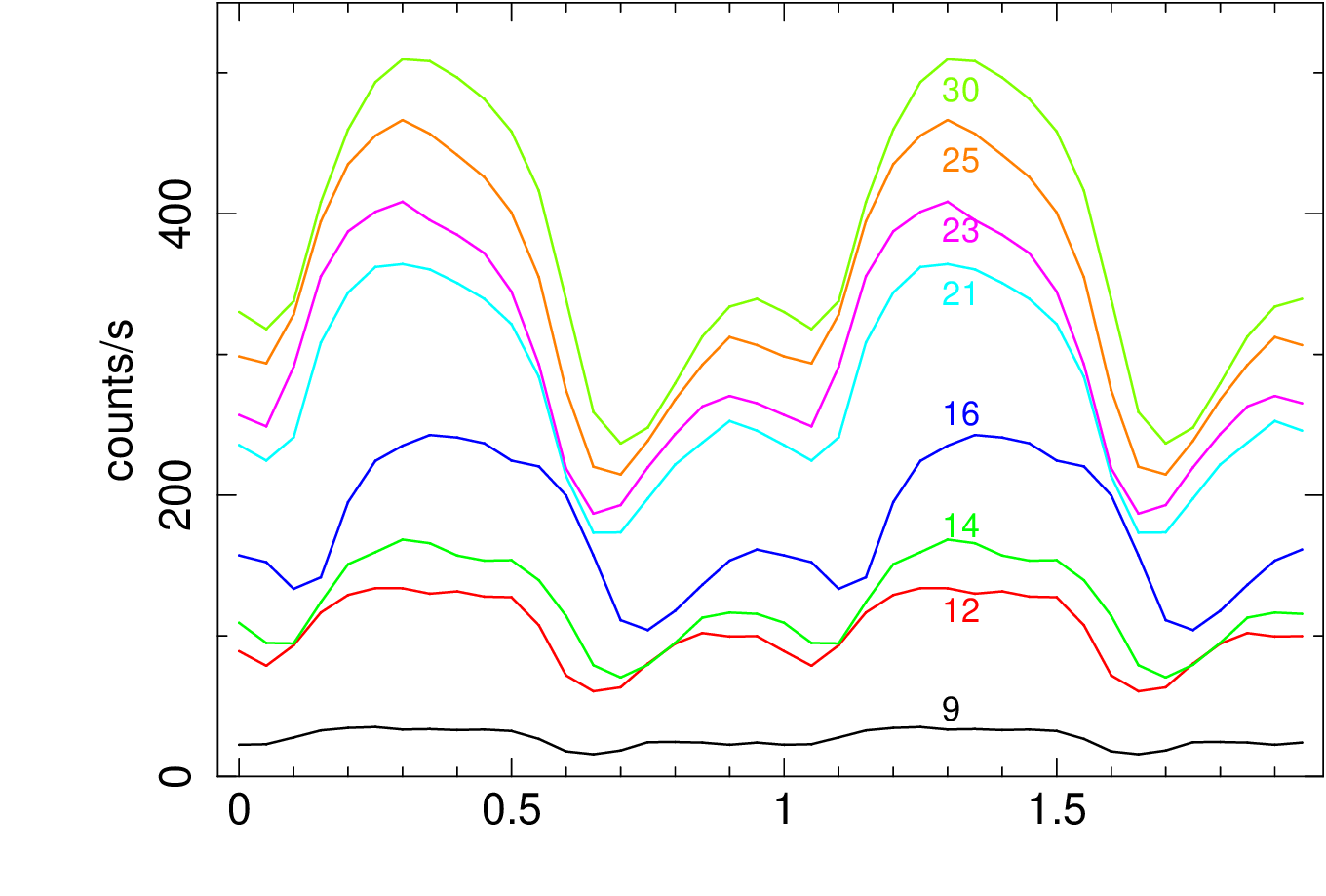}
	\vspace{-0.1cm}
	\includegraphics[width=3.4in]{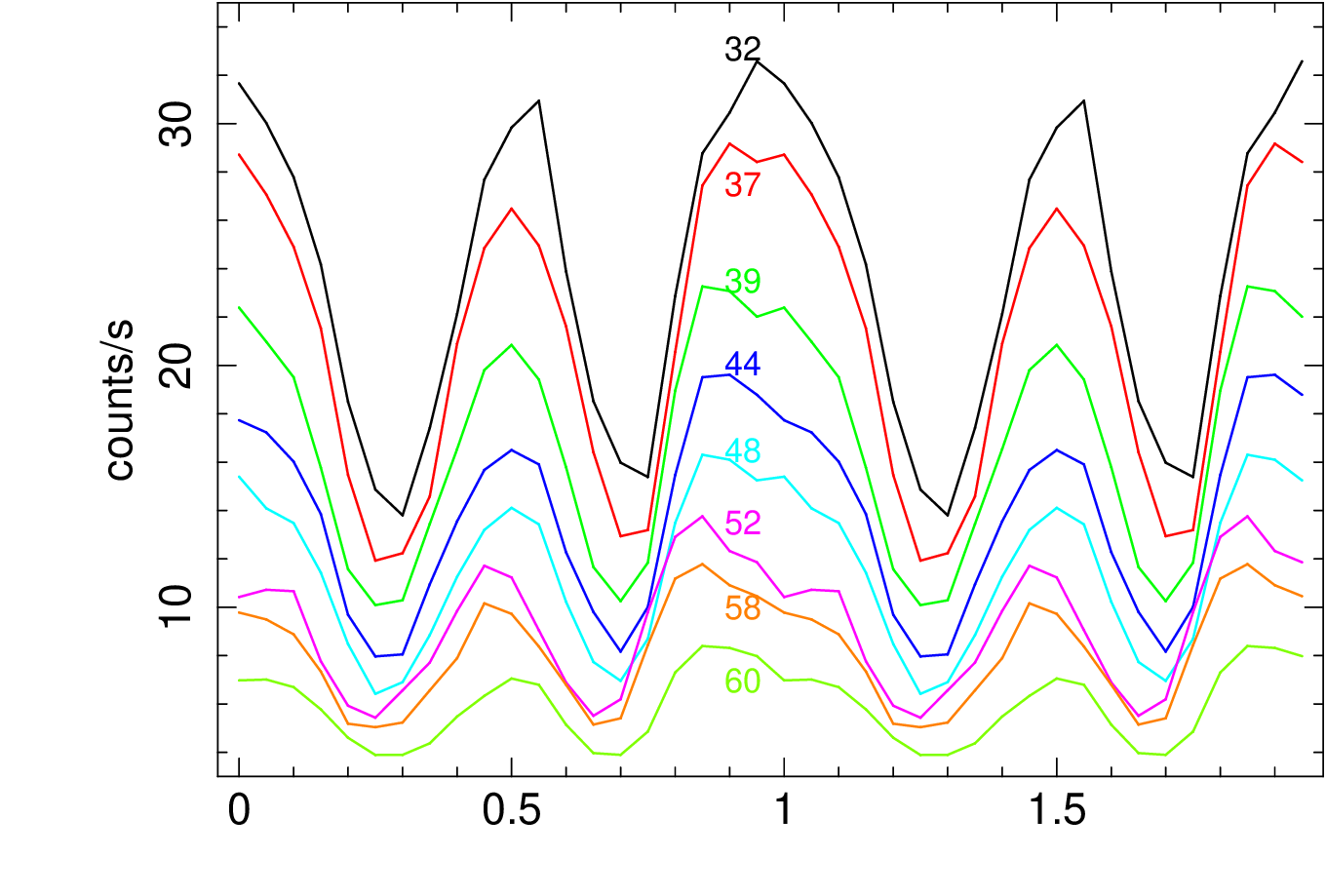}
	\includegraphics[width=3.4in]{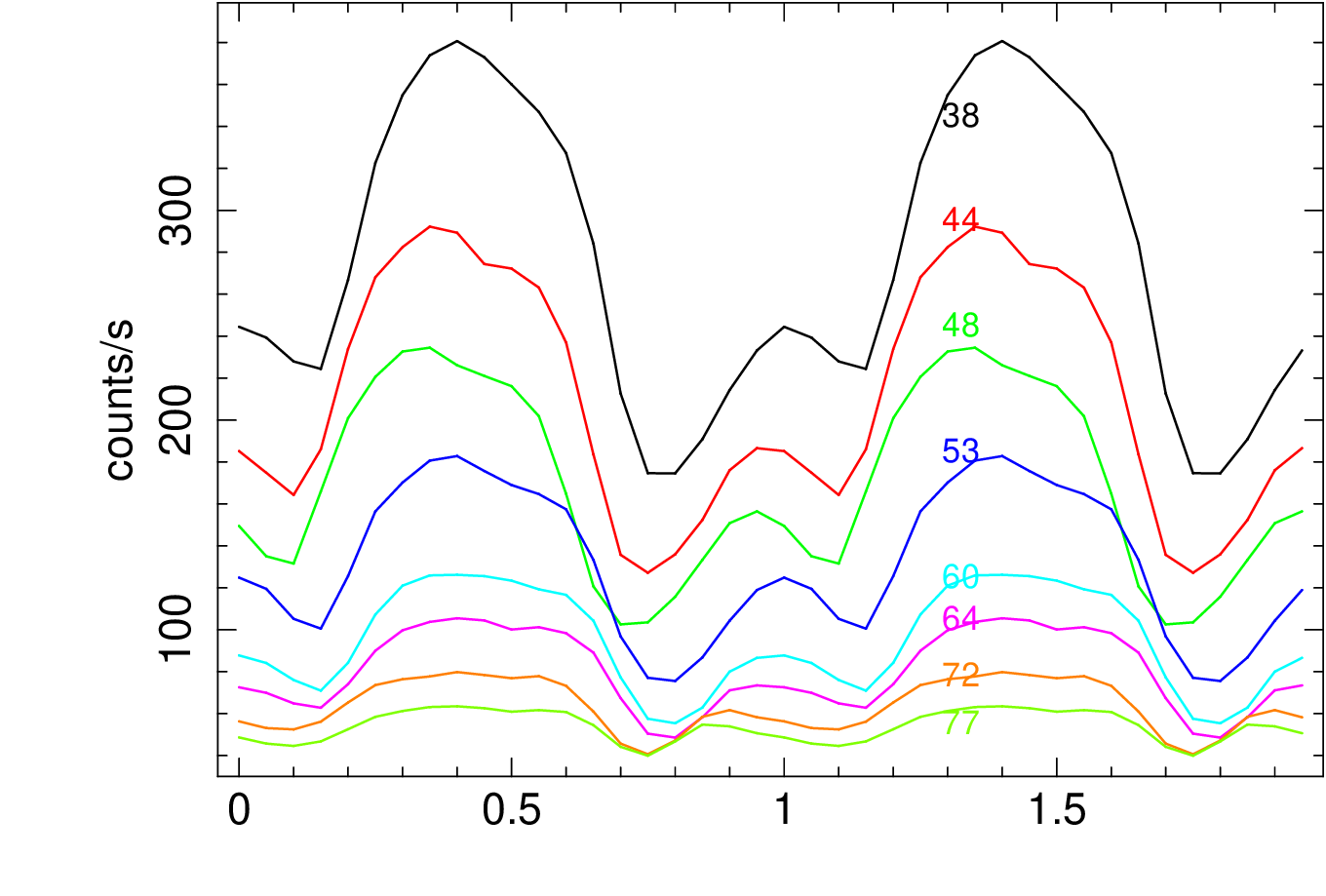}
	\vspace{-0.3cm}
	\includegraphics[width=3.4in]{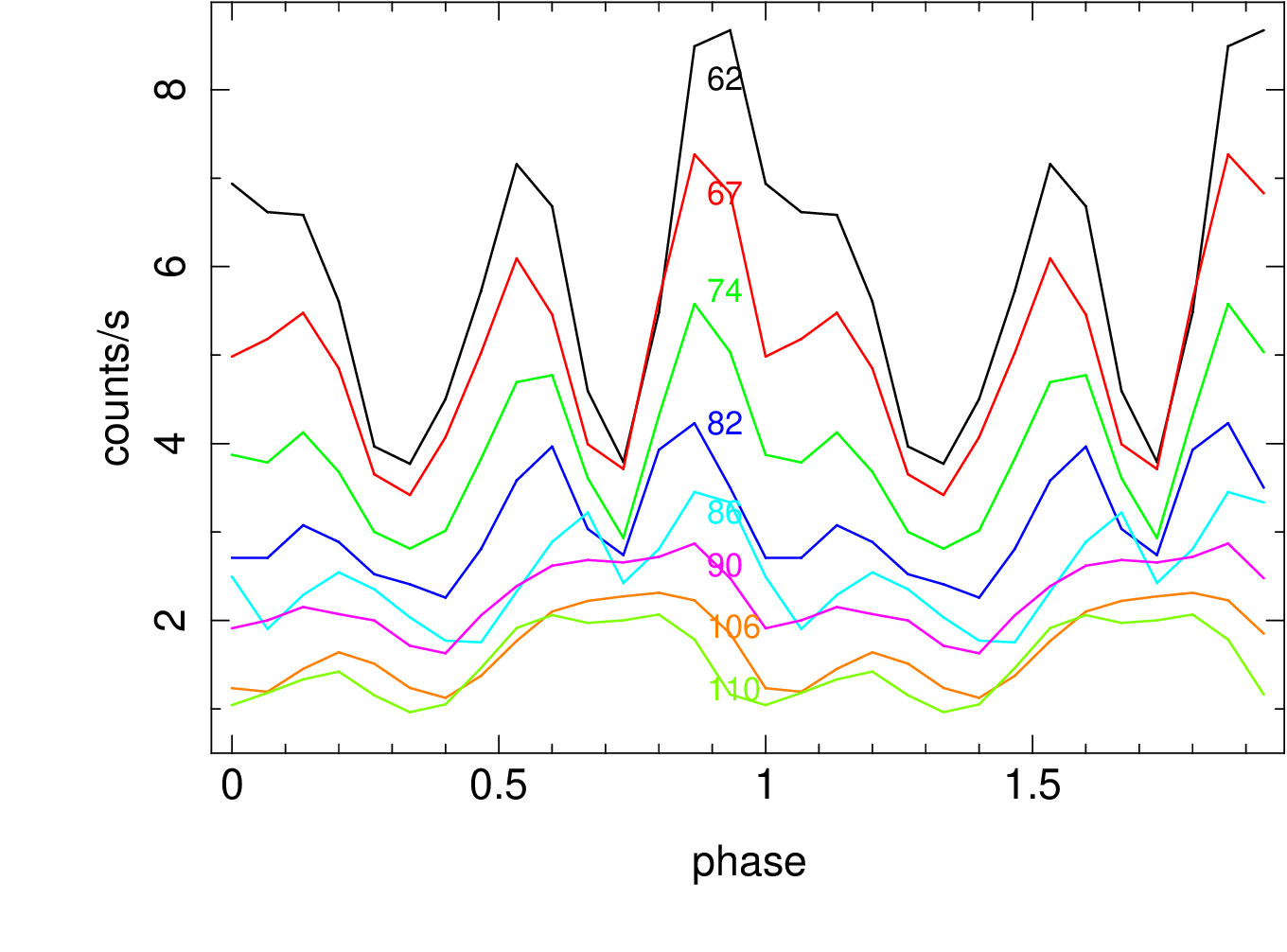}
	\includegraphics[width=3.4in]{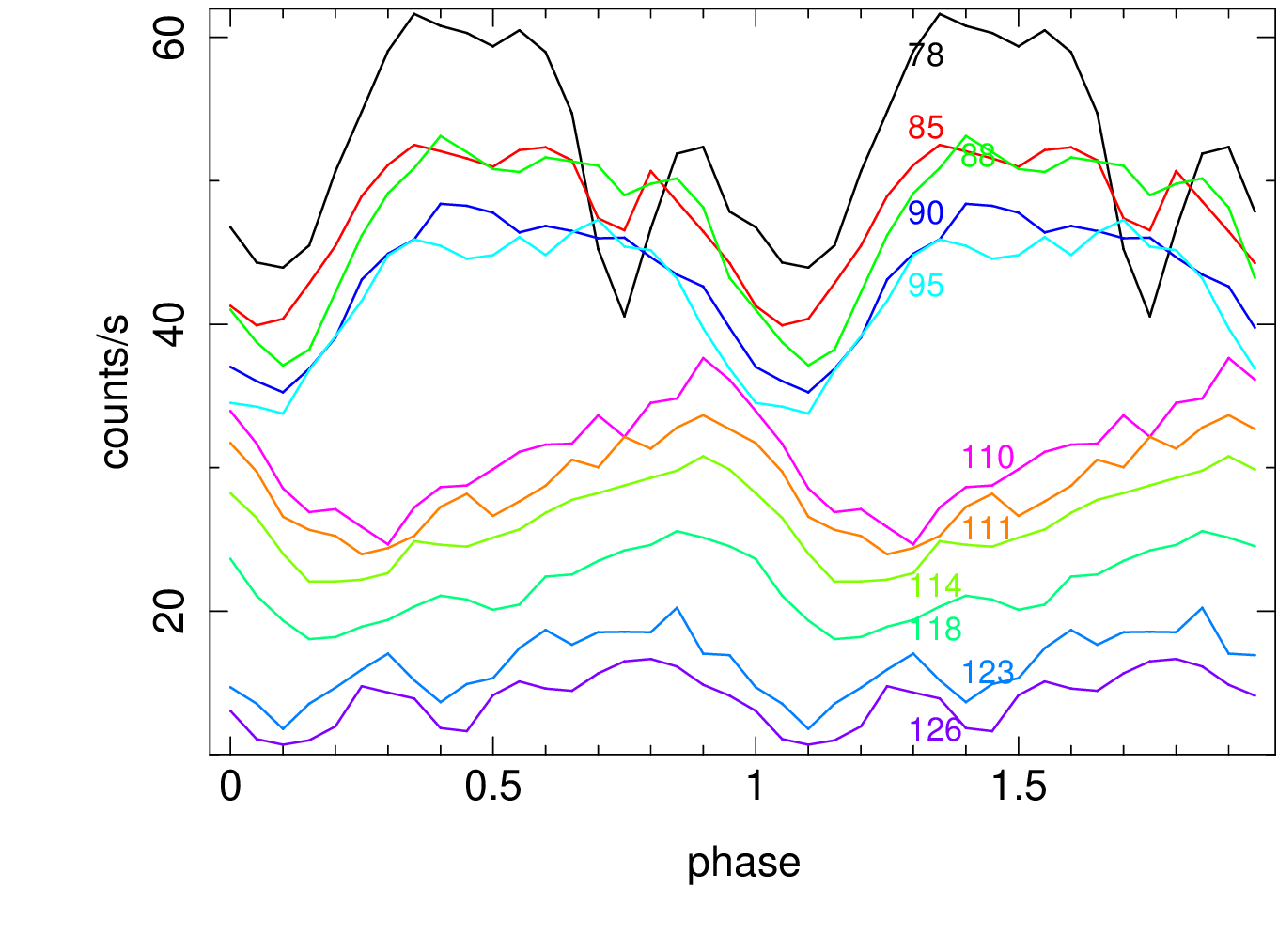}
	\caption{Time evolution of the Swift/XRT pulse profile of SMC X-3 
during its 2016 giant outburst (left) and the \nic profile of RX J0209
during its 2019 outburst (right).
The observed date of the profile since MJD 57600 and MJD 58800 
are marked for SMC X-3 and RX J0209, respectively.
The high double peaks of SMC X-3 (before MJD 57650) are 0.25 phase offset from 
	the low double peaks (after MJD 57700),
	indicating a 90 degree change of emission pattern.
	} 
\end{figure*}

\subsection{RX J0209}

We did a similar fitting to the observed spin frequencies of RX J0209
measured with \nic data \citep{Hou22}. 
The \nus observation of RX J0209 on MJD 58813 provides 
a conversion factor (BAT rate to flux) 
$f_1=1.01\times10^{-7}$ erg/counts for a BAT rate of $C_1=0.01$ counts\,cm$^{-2}$\,s$^{-1}$.
The \HXMT observations near the outburst peak around MJD 58828
provides $f_2=1.1\times10^{-7}$ erg/counts for a BAT rate of $C_2=0.016$ counts\,cm$^{-2}$\,s$^{-1}$. 
The differences of $f_1$ and $f_2$
reflect the spectral softening above 10 keV at higher fluxes, which was 
also found for Swift J0243.
So we adopt a conversion factor 
linearly interpolated over BAT rate $C$: $\frac{f_2-f_1}{C_2-C_1}(C-C_1)+f_1$.
As the public BAT light curve of RX J0209 
only started after MJD 58818, we use a BAT light curve kindly produced
by Amy Lien. We note that the bolometric correction is different from 
that in \citet{Hou22}, who did not use the BAT light curve.
To estimate the \nic fluxes, we used the {\it nibackgen3C50} tool 
to extract background and source spectra.
The corresponding light curves are plotted in the 
right panel of Figure 1. The 0.4-10 keV fluxes of RX J0209 do not exceed the level 
of the 10-70 keV fluxes even during the outburst peak. 

The fitting time range for RX J0209 is limited to MJD 58808-58890.
The fitting results are presented in the right panel of Figure 2 and 
listed in Table 1. Compared with the orbital parameters
by Fermi/GBM pulsar team, we obtain a smaller asin$i$
and also a larger eccentricity, similar to that of SMC X-3.
The fitted torque model index of $\alpha=0.97\pm0.02$ is 
also larger than the standard value of 6/7.

\section{Evolution of the pulse profile}

\subsection{SMC X-3}
The XRT pulse profiles of SMC X-3 had been presented in \citet{Weng17},
in a normalized sense, which is not sensitive to the growth/decline detail
of the absolute fluxes. Here we present a
phase-coherent XRT profile study, as did for Swift J0243 in \citet{Liu22}.
We assign a pulse phase $\phi(t)$ to the event time $t$ as 
$\phi(t)=\int_{t_0}^{t}\nu(t)dt$,
where $\nu(t)$ is the pulse frequency interpolated from measurements using 
a cubic spline function. 
The event arrival time $t$ is barycenter corrected and binary corrected 
with the orbital parameters obtained in previous section. 
The GBM-measured spin frequencies showed some fluctuations before 
MJD 57616, so we use the GBM-measured spin frequencies
within MJD 57616-57673 and XRT-measured values in other time.
On MJD 57618, 57634, 57664, and 57667, the obtained pulse profiles 
show apparent phase offsets, and we modify the measured spin frequencies 
a little bit around these times to make the profiles aligned.
The resulted pulse profiles within 0.6-10 keV are plotted in Figure 3. 

As can be seen, during the peak period of the outburst, the profiles 
are dominated by two peaks around phase 1 and 0.5. After MJD 57652
(the second panel), the main peak becomes more left-tilted, and the 
phase separation between the two peaks is about 0.35.
Around MJD 57667 (the third panel), 
a minor peak appeared about 0.25 phase right to the main peak. 
Around MJD 57680-57682, the phase separation of the original two peaks 
reduced to about 0.25.
Finally, around MJD 57690, 
the original two peaks merged together and 
became a broad bump around phase 0.7, and the profiles looked double-peaked again.
Note that the two low peaks after MJD 57690 are about 0.25 phase offset from those
during the outburst peak period, a phenomena just as that for Swift J0243.

\begin{figure*}
	\hspace{-0.2in}
	\includegraphics[width=3.4in]{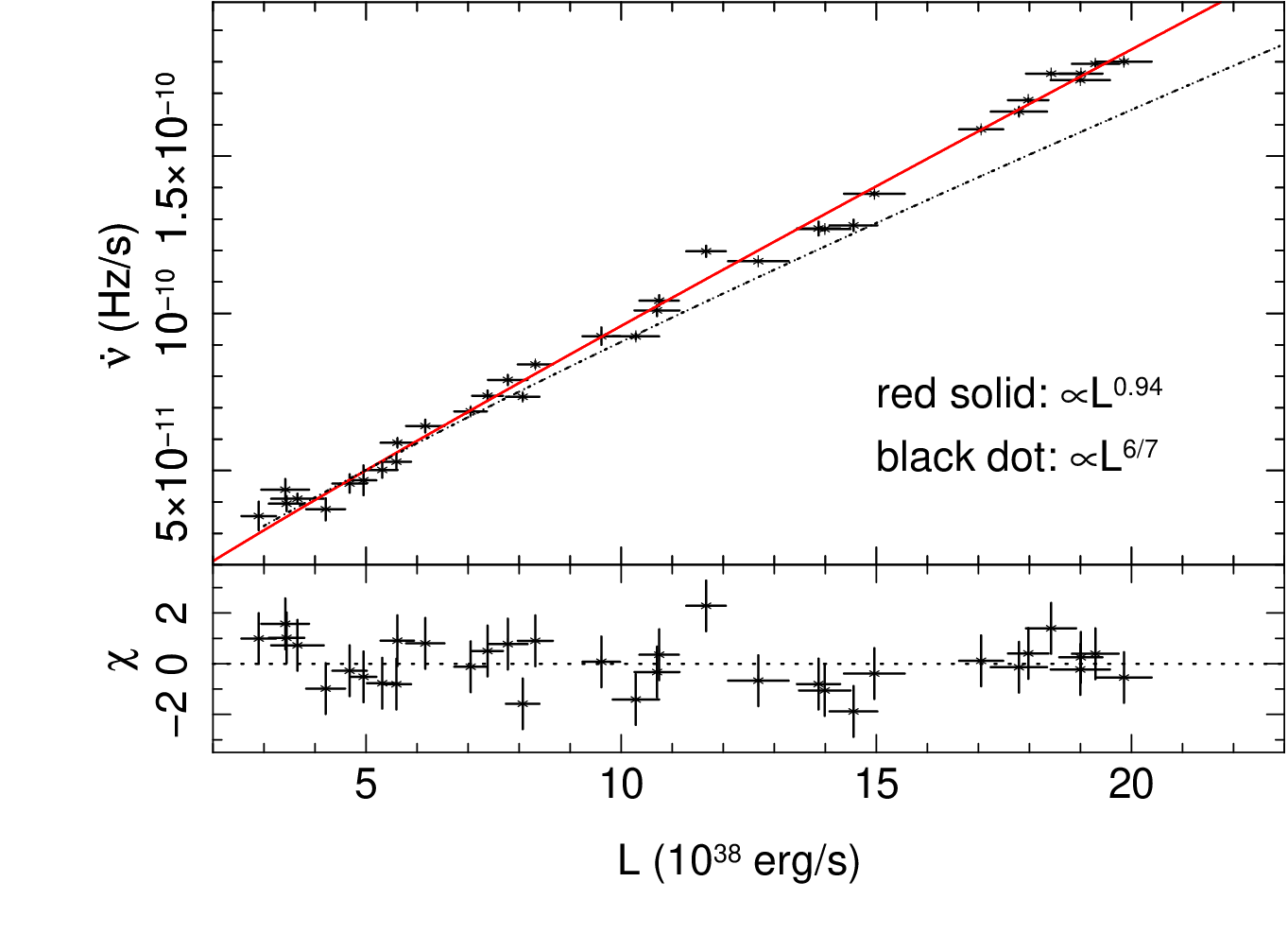}
	\includegraphics[width=3.4in]{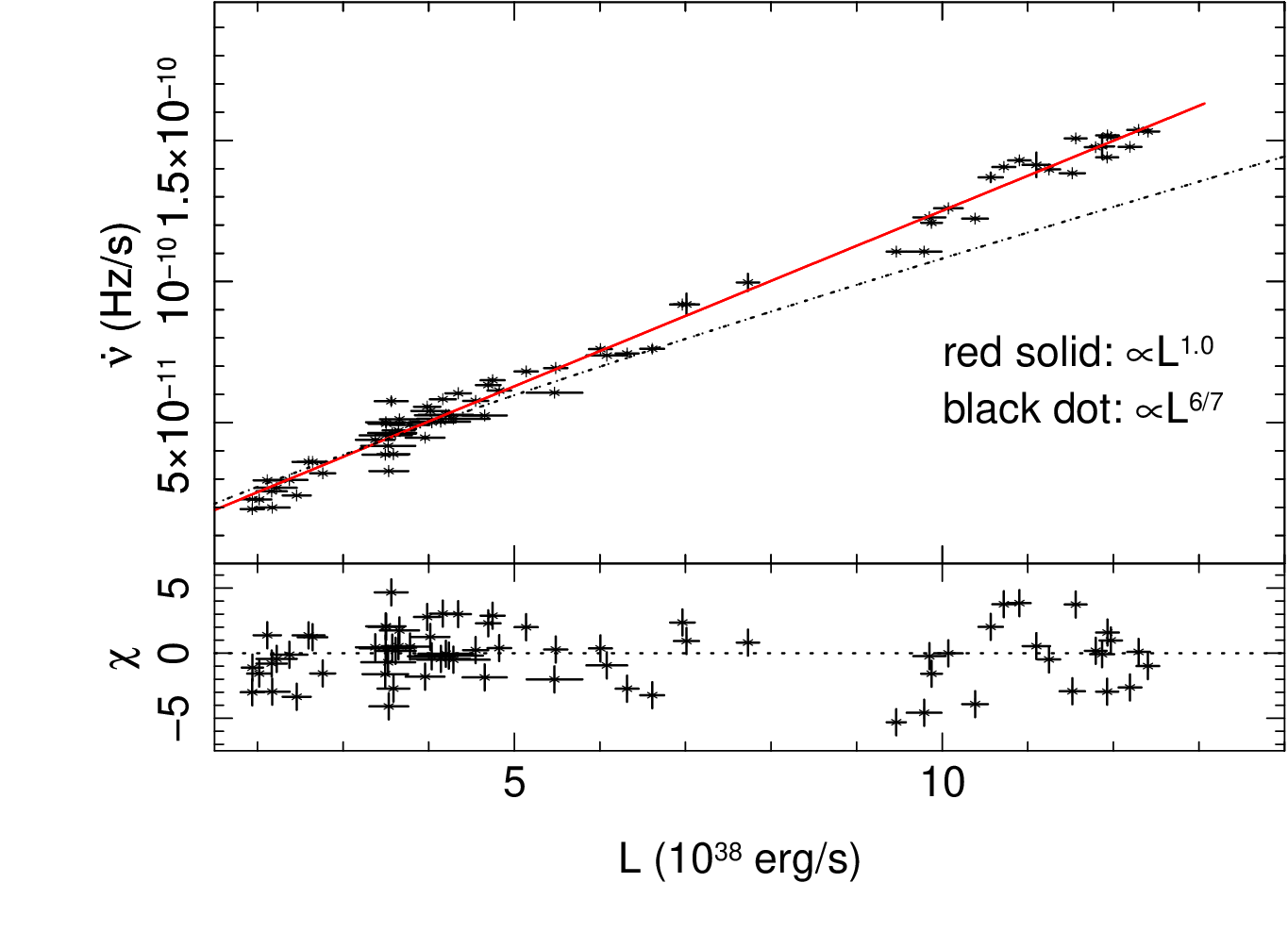}
	\caption{
Spin-up rate vs bolometric flux for SMC X-3 (left) and RX J0209 (right).
 The data can be fitted
	with a power-law of $\dot{\nu}\propto L_{bol}^{0.94\pm0.03}$ 
and $\dot{\nu}\propto L_{bol}^{1.00\pm0.03}$ (the red solid line), respectively.
The classical relation of $\dot{\nu}\propto L^{6/7}$
is also plotted (the black dot line). 
}
\end{figure*}

\begin{figure}
	\hspace{-0.2in}
	\includegraphics[width=3.5in]{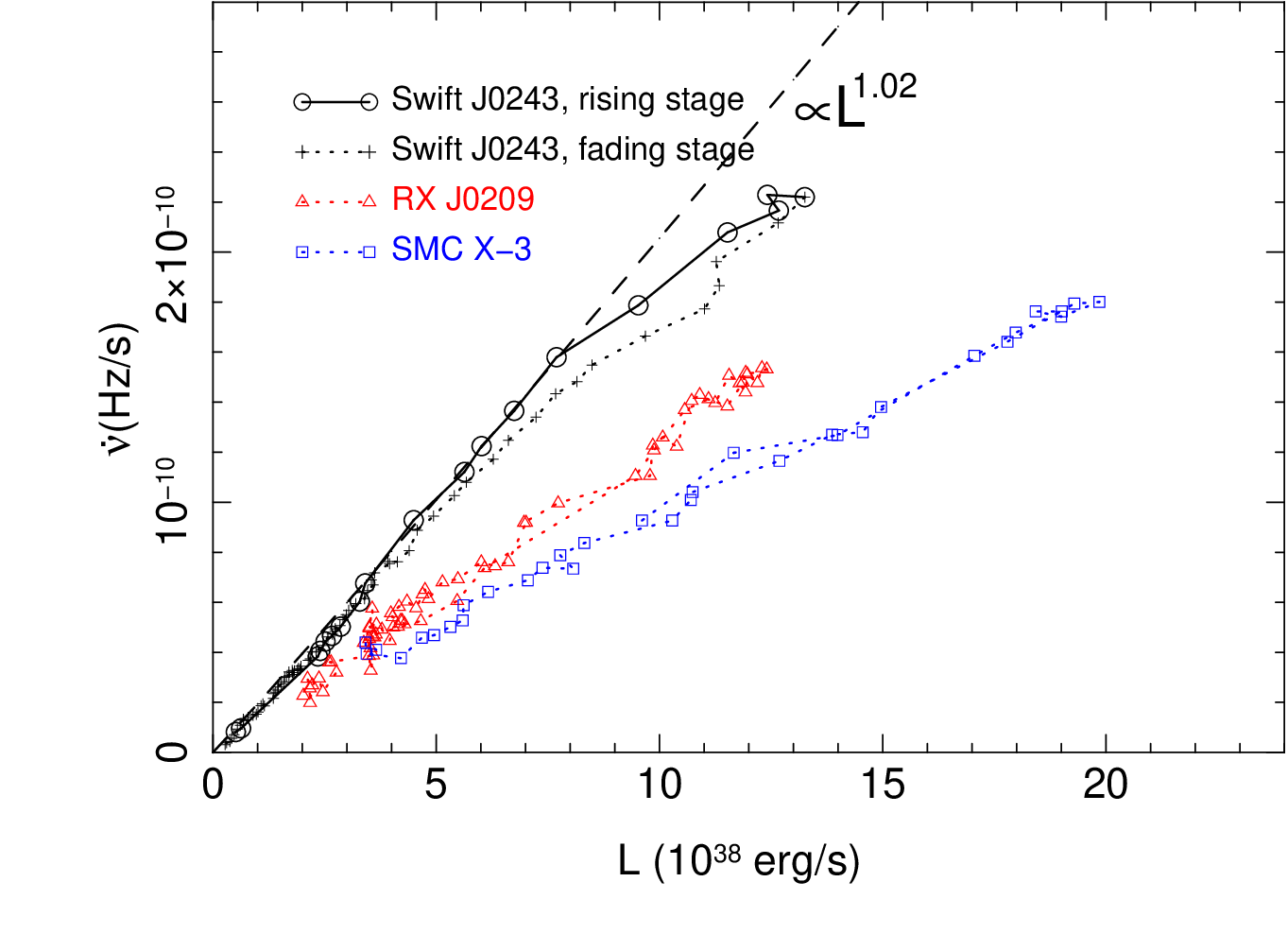}
	\caption{
Spin-up rate vs bolometric flux for Swift J0243, SMC X-3, and RX J0209.
}
\end{figure}

Taken the high double peaks as from a fan beam, the low double peaks would be 
from a pencil beam. The transition of the fan beam to the pencil beam, 
that is, the disappearance of accretion column, happened around MJD 57686-57690, 
when the two peaks of the fan beam merged into one.
The corresponding luminosity
is around $2.3\times10^{38}$erg\,s$^{-1}$.
The high double peaks (the fan beam) take their full shape before 
MJD 57662, corresponding to a luminosity of $6\times10^{38}$erg\,s$^{-1}$.
In between, the fan beam gradually changes to the pencil beam.

\subsection{RX J0209}

The corresponding \nic pulse profiles within 0.4-10 keV of RX J0209 are presented in the 
right panels of Figure 3. (The \nic profile of RX J0209 was studied 
in a normalized sense in \citet{Hou22}.) During the outburst peak, the profiles are also 
composed of two peaks, although the minor peak is relatively fainter compared 
with the main peak. Around MJD 58871-58872 (the second panel), the previous left-tilted 
main peak becomes more symmetric, and the minor peak becomes 
more left-titled. Around MJD 58885-58888, the minor
peak merged with the main peak and the profile became a broad bump.

Due to the lack of data within MJD 58896-58909, we can not align well the pulse 
profiles after MJD 58909 to those before MJD 58896. Nevertheless, it is clear 
that the broad bump on MJD 58895 evolved to an right-tilted peak around MJD 58910. 
A likely evolution path is that the intensity of the left part 
of the bump (the main peak of the high double peak) decreased,  
and that produced a right-tilted peak, if assuming the valley phase of the 
profile is not changed much within MJD 58896-58909. 
This would make the peak on MJD 58910 
located near the valley of the high double peak.
The profiles show two peaks again after MJD 58923. 
We note that there seems already a feature of minor peak around MJD 58911-58918, 
although not as apparent as those after MJD 58923.
Thus, we can see that 
the high double peaks changed to one bump around MJD 58888, and then 
changed to low double peaks after MJD 58911, which is also 
likely 0.25 phase 
offset from the high double peaks. The corresponding 
critical luminosity (around MJD 58911-58918) is about $1.25\times10^{38}$erg\,s$^{-1}$.

\section{The $\dot{\nu}$-$L$ relation}

For the super-Eddington regime of Swift J0243, we found a linear relation
of $\dot{\nu}\propto L$ within $5-13\times10^{38}$erg\,s$^{-1}$. 
A linear-like $\dot{\nu}$-$L$ relation is also implied during the fitting process 
of the orbital parameters of SMC X-3 and RX J0209.
To obtain the $\dot{\nu}$-$L$ relation of SMC X-3, we convert the GBM 
measured spin frequencies with the fitted orbital parameters listed in Table 1.
We calculate an averaged bolometric fluxes between two successive XRT observations, 
and calculate $\dot{\nu}$ at the intermediate times. 
The calculated spin-up rate
against the bolometric flux is plotted in Figure 4. 
The data is limited to MJD 57681, beyond which the high two 
peaks started to merge together and the inferred $\dot{\nu}$ showed 
larger fluctuations.

The $\dot{\nu}$-$L$ relation of SMC X-3 can be fitted with a 
power-law model of $\dot{\nu}\propto L_{bol}^{0.94\pm0.03}$, consistent with 
that obtained during the fitting process of the orbital parameters.
On the other hand, the $\dot{\nu}-L$ relation of Swift J0243 shows a 
flattening at the highest luminosity.
The $\dot{\nu}-L$ relation of SMC X-3 shows no such a flattening.

The $\dot{\nu}$-$L$ relation of RX J0209 is presented in the right panel of Figure 4.
It can be fitted with a power-law model of 
$\dot{\nu}\propto L_{bol}^{1.00\pm0.03}$.
The maximum luminosity of RX J0209 is about 
$1.2\times10^{39}$erg\,s$^{-1}$, 
about half of that of SMC X-3.

To compare the $\dot{\nu}$-$L$ relation for all three sources, the observed 
$\dot{\nu}$-$L$ data for Swift J0243, SMC X-3, and RX J0209 are plotted together 
in Figure 5.
The luminosity of Swift J0243 is within 0.5-150 keV band, 
a little broader than the 70 keV limit for SMC X-3 and RX J0209.
We note that the previous estimation of the luminosity of Swift J0243
assumed a distance of 6.8 kpc from {\it Gaia} data release 2 \citep{Bai18}, 
while the {\it Gaia} early data release 3 provided a distance 
of 5.2 kpc for Swift J0243 \citep{Bai21}. If taking this new distance, the luminosity 
of Swift J0243 would be 0.6 times smaller than previous estimation. 
We have applied this reduction of the luminosity of Swift J0243 in Figure 5.
(If not applying this reduction, the data
points of Swift J0243 will locate among
those of RX J0209.)
As can be seen, 
the data points of SMC X-3 are lying below those of RX J0209, 
which are below those of Swift J0243.
That is, for a given observed luminosity, the spin-up rate of the 
neutron star of Swift J0243 is higher than that of SMC X-3 and RX J0209.

\section{Discussion and conclusion}

We analyzed the 2016 giant outburst of SMC X-3 and 2019 giant outburst of 
RX J0209, with the aim to compare their super-Eddington accretion 
regime with that of Swift J0243 in 2017. 
To find the orbital parameters of SMC X-3 and RX J0209, 
we estimated their bolometric fluxes based on Swift/BAT fluxes in 15-50 keV
and Swift/XRT and \nic fluxes below 10 keV. 
We have limited the fitting time range of observed spin frequencies
to the outburst period, as the accretion torque is dependent 
on the accretion luminosity.
The high double-peak profile of SMC X-3 
is found to be 0.25 phase offset from that below $2.3\times10^{38}$erg\,s$^{-1}$. 
The profile of RX J0243 shows a similar high double-peak to 
low double-peak transition around $1.25\times10^{38}$erg\,s$^{-1}$,
although the profile phase at low luminosity is not well aligned.
These pulse profile behaviors are just the same as that for Swift J0243, 
the high double peak of which showed a 0.25 phase offset from 
that below $0.9\times10^{38}$erg\,s$^{-1}$,
if assuming a distance of 5.2 kpc for Swift J0243.

In the standard accretion scenario of magnetized neutron star 
\citep[e.g.][]{GS73,Dav73,BS76}, 
at low accretion rates, X-ray radiation is emitted in a
pencil beam along the magnetic pole; above a critical luminosity,
an accretion column formed below a radiation-dominated shock, 
and X-ray photons are emitted mainly from the sidewall of the column in a fan
beam. The angle difference betwwen a pencil beam and a fan beam is 90 degree.
Therefore, the 0.25 phase offset between the high and low double-peak profiles, 
which corresponds to a 90 degree change of the emission beam,
is a strong evidence for a transition from a high fan beam to a low pencil beam
and for the existence of accretion column during the high double-peak regime.
The critical luminosity depends on the accretion geometry and the magnetic field.
Above $\sim10^{37}$ erg/s, the critical luminosity grows with the magnetic 
field because of 
the drop of the effective cross-section below the cyclotron energy. It 
roughly increases linearly with the filed strength \citep{Mus15,Bec12}.
Recently, a measurement of cyclotron line around 130 keV was reported 
during the main pulse at the outburst peak of Swift J0243 with \HXMT observations, 
which corresponds to a surface field above $1.6\times10^{13}$ G
\citep{Kong22}.
Taken it as the face value and a distance of 5.2 kpc for Swift J0243, the surface
magnetic field strength 
of SMC X-3 and RX J0209 would be around $4\times10^{13}$ and $2\times10^{13}$ G, respectively.

Above the profile transition (critical) luminosity, the profiles are not dominated by the fan beam immediately, 
but there is an intermediate stage for the fan beam to grow up and to dominate the emission pattern.
During this intermediate stage, the profile could look like one-peaked (Swift J0243), 
broad-bump (RX J0243), or three-peaked (SMC X-3) shape. Such intermediate profile shapes 
would depend on our viewing angle towards the emission beam and the bending of light near 
the neutron star, and may also depend on the magnetic field. Detailed investigations are 
needed to see what kind of constraints these intermediate profiles may provide.
In principle, the pulse profile could also be affected by the growth of accretion column,
and a sudden change may be expected when the column exceeds a height and 
the column can not be eclipsed by the neutron star anymore \citep{Mus18}.
Nevertheless, the predicted dip structure of the eclipse is not observed for 
both SMC X-3 and RX J0209, and there also seems no a characteristic 0.25 phase offset
that was expected by a sudden appearance of eclipsed column.

The observed $\dot{\nu}-L$ relation of SMC X-3 follows as 
$\dot{\nu}\propto L^{0.94}$, 
while that of RX J0209 follows as $\dot{\nu}\propto L^{1.00}$. 
Such a relation is similar to that of Swift J0243 within 
$3-8\times10^{38}$erg\,s$^{-1}$. 
In the RPD regime as modeled by \citet{Cha19,Cha17}, 
the magnetosphere radius is almost independent of the accretion 
rate. As a result, a linear-like relation
of $\dot{\nu}\propto L$ is predicted, which is consistent with 
the observed $\dot{\nu}-L$ relation. Note that a larger 
magnetosphere radius is expected for a stronger field: $R_{in}\propto\mu^{4/9}$
in the model of \citet{Cha19}. If this is true, one would expect a larger 
$\dot{\nu}$ at similar luminosity ($2\pi I\dot{\nu}=\dot{m}\sqrt{GMR_{in}}$, neglecting
the braking torque) for a stronger magnetic field. In contrast, the observed $\dot{\nu}$ of 
SMC X-3, with the highest critical luminosity (magnetic field), lies below those of Swift J0243 and 
RX J0209.

One possibility is that the dipole component of SMC X-3 is much weaker than 
the multi-pole component in the vicinity of the neutron star, as proposed by \citet{Tsy17}.
Such an idea has also been proposed for other ULXPs \citep{Isr17a,Isr17b},
and also for Swift J0243 \citep{Kong22} and RX J0209 \citep{Hou22}. 
If this is true, SMC X-3 would have the 
largest multi-pole field, but with a smallest accretion torque among all three sources.
We note that a multipolar magnetic structure might increase the accreting 
area of the neutron star and affect the critical luminosity as discussed recently 
for Her X-1 \citep{Mon22}.

On the other hand, we noted that the co-rotation radius of SMC X-3 is 
about $6.7\times10^8$ cm, which is comparable
with the predicted magnetosphere 
radius ($2\times10^8$ cm) of a source with a magnetic moment $\sim2\times10^{31}$G\,cm$^3$ around 
$10^{39}$erg\,s$^{-1}$ \citep{Cha19} (their Figure 13).
If we take the braking torque as on the order of $\mu^2/R_c^3$
\citep[e.g.][]{Lip92}, it is indeed similar to the accreting torque
with a luminosity of $10^{39}$erg\,s$^{-1}$.
Thus, one needs to take into account the braking torque in the situation of SMC X-3.
\citet{Cha19} has assumed that the disk-magnetosphere interaction 
is confined to a narrow region near the inner radius of the disk.
Adopting a wide interaction/threading region might modify the prediction 
of a RPD disk, and further modelling is needed to check whether a smaller torque 
would be resulted for accretion through a RPD disk with a strong dipolar field.

Above $8\times10^{38}$erg\,s$^{-1}$ (with a distance of 5.2 kpc), the minor peak of 
Swift J0243 showed a reduced intensity compared with the main peak, and 
the $\dot{\nu}-L$ relation becomes flattened \citep{Liu22}. 
The peak luminosity of RX J0209 
reached a level of $1.2\times10^{39}$erg\,s$^{-1}$, 
while that of SMC X-3 reached $2\times10^{39}$erg\,s$^{-1}$. 
It is interesting to see that neither flattening of $\dot{\nu}-L$
relation, nor
the reduction of faint peak were observed for SMC X-3 and RX J0209. 
On the one hand, this is consistent with the assumption that these two 
phenomena of Swift J0243 are connected
somehow, like through irradiation of the accretion disk by the accretion 
column \citep{Liu22}.
On the other hand, 
it indicates no apparent irradiation (feedback) from the accretion column 
of SMC X-3 at its highest luminosity. In principle, the irradiation of the 
disk depends on the geometry of the accretion column relative 
to the accretion disk, which is hardly constrained. 
If taking $R_{in}\propto\mu^{4/9}$, the magnetosphere radius 
of SMC X-3 during the RPD regime is estimated to 
be about 1.5 times larger than that of Swift J0243. This will make the irradiation 
flux per unit area on the inner region of the disk relatively fainter
for SMC X-3, 
reduced by a factor $\sim2.3$.
If the disk and column configuration of SMC X-3 
is similar to that of Swift J0243, one would expect some changing 
of the pulse profile and $\dot{\nu}-L$ relation above $2.3\times8\times10^{38}=1.84\times10^{39}$erg\,s$^{-1}$, 
which is just reached by the peak luminosity of SMC X-3. It seems a luminosity 
higher than $2\times10^{39}$erg\,s$^{-1}$ is needed to probe the irradiation
effect in SMC X-3. While for RX J0209, a luminosity higher than 
$1.2\times10^{39}$erg\,s$^{-1}$ is needed.

We note that there are some uncertainties not accounted for when estimating 
the observed $\dot{\nu}-L$ relation. For example, we have included all the 
soft band fluxes when calculating the bolometric luminosity. In reality, 
part of the soft emission could come from the accretion disk (although the fraction
should be small as the disk is truncated far away),
and one may only use the luminosity from the accretion column to estimate 
mass accretion rate. This would likely make the index of $\dot{\nu}-L$ relation
a little larger. For sources of relatively lower luminosity
(below $10^{38}$erg\,s$^{-1}$), indexes of 
$\dot{\nu}-L$ relation around 1 had been reported in the literature
\citep[e.g.][]{Bil97,Fil17}. If this is true, it would be interesting 
to compare the origin of linear-like $\dot{\nu}-L$ relation below 
and above Eddington luminosity. We noted that the bolometric correction is critical to estimate the index, 
and a broadband spectroscopy is necessary to check and calibrate the correction.
The intense broadband monitoring of Galactic X-ray pulsars by \HXMT provides a chance to 
study these less-luminous sources, and we plan to study them in a future work.

In summary, the high double peaks of SMC X-3
are found to be 0.25 phase offset from the low double peaks, just as that for Swift J0243, and It seems also the case for RX J0209. 
These results support for the formation of an accretion column during the high luminosity state in these 
three sources. Among them, SMC X-3
has a highest critical luminosity around 
$2.3\times10^{38}$erg\,s$^{-1}$, which leads 
to an estimation of a highest magnetic field $\sim4\times10^{13}$ G. 
Such a high magnetic field is strong enough to support electron-position annihilation 
and may produce a neutrino pulsar \citep{Mus19}. 
These results indicate that the population of ULXPs likely corresponds to X-ray pulsars of highest magnetic field. The $\dot{\nu}-L$ relation of all 
three sources follows a linear-like relation,
consistent with accretion through a RPD disk.
Among all sources, only Swift J0243 showed a 
flattening of $\dot{\nu}-L$ relation and reduction of the faint peak at the 
highest luminosity. Nevertheless, it does not mean these anomalous behaviors are unique to Swift J0243, because 
the peak luminosity of SMC X-3 and 
RX J0209 may be not high enough to probe this behavior. One should be cautionary when 
dealing with the $\dot{\nu}-L$ relation of ULXPs.

\section*{Acknowledgements}
We thank our referee for helpful comments, Amy Lien for help of 
producing a BAT light curve of RX J0209.
This work made use of data from 
\sw, \fm, {\it Nicer}, and \HXMT. We acknowledge the support by National 
Natural Science Foundation of China (U1938113, U1938103, U2038103, 11733009, 12173103 and U2038101),
and the Scholar Program of Beijing Academy of Science and Technology (DZ BS202002). 

\section*{Data Availability}
The data underlying this article are publicly available at
https://swift.gsfc.nasa.gov/, https://heasarc.gsfc.nasa.gov/docs/nicer/index.html,
and http://archive.hxmt.cn/.

\bibliographystyle{mn2e}

\appendix

\end{document}